\documentclass[a4paper,12pt]{article}
\usepackage{graphicx}
\usepackage{amsbsy}
\usepackage{amsbsy}
\usepackage{amsmath}
\usepackage{amssymb}
\usepackage{amsfonts}

\def\r#1{_{\rm #1}}
    \def\p#1#2{\frac{\partial{#1}}{\partial{#2}}}

\def\Mo{M_\odot}
\def\Lo{L_\odot}

\def\mec#1#2{m_{\rm e}^{#1}c^{#2}}

\def\ee{{\rm e}}

\def\MM#1{\left(\frac{M}{M_\odot}\right)^{#1}}
\def\TK#1{\left(\frac{T_\infty}{10^4~{\rm K}}\right)^{#1}}
\def\rhoinf{\left(\frac{\rho_\infty}{10^{-24}~{\rm g~cm^{-3}}}\right)}

\begin{document}

\begin{center}
{ \Large
\textbf{Accretion into black hole, and formation of magnetically
arrested accretion disks}
}\\[20pt]

{ \large

G.S. Bisnovatyi-Kogan$^{1,2,3}$
}\\[20pt]

$^{1}$Space Research Institute of Russian Academy of Sciences,\\ Profsoyuznaya 84/32, Moscow 117997, Russia;\\
$^{2}$National Research Nuclear University MEPhI (Moscow Engineering\\Physics Institute), Kashirskoe Shosse 31, Moscow 115409, Russia\\
$^{3}$ \quad Moscow Institute of Physics and Technology MIPT, \\Institutskiy Pereulok, 9, Dolgoprudny, Moscow region, 141701.\\[20pt]

email: gkogan@iki.rssi.ru\\[20pt]


\end{center}

\abstract{The exact time-dependent solution is obtained for a magnetic field growth during a spherically symmetric accretion  into a black hole (BH) with a Schwarzschild metric.
Magnetic field is increasing with time, changing from the initially uniform into a quasi-radial field.
Equipartition between   magnetic and kinetic energies in the falling gas is supposed to be established in the developed stages of the flow. Estimates of  the  synchrotron radiation intensity are presented for the stationary flow. The main part of the radiation is formed in the relativistic region $r \leq 7 r_g$, where $r_g$ is a BH gravitational radius.
The two-dimensional stationary  self-similar magnetohydrodynamic solution is obtained for the matter accretion into BH, in a presence of a large-scale magnetic field, under assumption, that the magnetic field far from the BH is homogeneous and its influence on the flow is negligible. At the symmetry plane perpendicular to the direction of the distant magnetic field, the dense quasi-stationary disk is formed around BH, which structure is determined by dissipation processes.  Solutions of the disk structure have been obtained for a laminar disk with the Coulomb resistivity, and for a turbulent disk.
Parameters of the shock forming due to matter infall onto the disk are obtained.
The radiation spectrum of the disk and the shock are obtained for the $10\,\, M_\odot$ BH. The luminosity of such object is about the solar one, for a characteristic galactic gas density, with possibility of observation at  distances less than 1 kpc. The spectra of a laminar and a turbulent disk structure around BH are very different.
The laminar disk radiates mainly in the ultraviolet, the turbulent disk emits a large part of its flux in the infrared. It may occur that some of the galactic infrared star-like sources are a single BH in the turbulent accretion state. The radiative efficiency of the magnetized disk is very high, reaching $\sim 0.5\,\dot M\,c^2$. This model of accretion was called recently as a magnetically arrested disk (MAD). Numerical simulations of MAD, and its appearance during accretion into neutron stars are considered and discussed.
}

\section{Introduction}

The aim of this paper is to present a review of works on the modeling of the accretion into a black hole (BH) of a gas with a large scale magnetic field. Earlier works on this topic made by the author \cite{bkr74,bkr76}, have been done using a simplified analytical solutions, along with approximate one-dimensional solutions. The results of these works have been in large confirmed by 2 and 3D numerical simulations, made recently by the members of the group  \cite{ ig08,tnm11,cnj18}, which revived the interest to this model under the name MAD. It follows from this review, that extensive numerical simulations  made by this group are still not powerful enough, to reproduce all features of the MAD model, which can be studied in the simplified analytical models. An example of it is a strong shock wave formation because of collision with a disk of the gas, falling along magnetic field lines.

\subsection{Black holes and accretion}

\noindent A BH is formed, if a mass of a collapsing core exceeds the mass limit of
the neutron star (NS). A BH is an object with a very strong
gravitational field $\varphi_G\sim c^2$, described by the general relativity.
 BH properties are studied in many
of monographs (see, for example, \cite{110,169,zn71}). The most important
observational property of a BH is that it does not permit
a light escape, so that in vacuum it could be detected only
by a black spot (shadow), on a background of twinkling sources (stars).
 The space between stars and galaxies is
 filled with a gas that may fall into BH, being heated.
The radiation emitted by this gas, could make a BH to become visible.
A powerful accretion, and best observational conditions occur when
the black hole is in a binary with a normal star, and the material flows from this
star on to BH. The observed X-ray sources in binaries--candidates
for black holes are: CygX-1, LMCX-1, A0620-00,
4U1658-48(GX339-4) and others \cite{222a,222b,222c,478}.

We consider here models for accretion onto BH of a magnetized matter. These
models are hydrodynamical, i.e., the mean free path of particles is assumed
to be less than the binary size. For  ionized interstellar medium,
this is connected with an entanglement of electron and ion trajectories by
the magnetic field, and in binaries, where the density is high, the
mean free path is determined by Coulomb collisions and is also short enough.

\subsection{Accretion of magnetized matter}

The magnetic fields play a very important and in some cases a critical
role in a possibility of observations of compact objects, NS
and BH. A BH may be observed only when matter
falls into them, producing radiation of a hot gas.

It was proposed to search for a BH in binaries, where radiation due to accretion may be large \cite{zn71}.
It was shown in \cite{shw71} that during a spherically symmetric accretion
  into a single black hole the
transformation of the kinetic energy into radiation is very small in the absence of
magnetic field, with the efficiency $\eta=L_{bh}/\dot M c^2$ of order $\sim 10^{ -8}$ \cite{shw71}, $L_{bh}$ is a luminosity, and $\dot M$ is a mass flux into BH per unit time.
Such an object cannot
be observed from the Earth. The situation is drastically changes for
 a magnetized matter falling into BH, when the efficiency strongly increases, and  luminosity becomes equal to $0.1\,\dot M\, c^2$  \cite{shw71}. The account of the magnetic field had been done in \cite{shw71}, for a completely irregular field. The equipartition between the kinetic energy of the falling matter, and the magnetic energy during the accretion, was assumed. The condition of equipartition was supported by annihilation of the magnetic field lines, or some other dissipation. Such a flow was considered  in the Newtonian approximation, with phenomenological account of general relativity by the cut-off of the solution at $r = 1.5\,r_g$, where $r_g = 2GM/c^2$ is the gravitational radius in Schwarzschild metrics.
The approximation of complete irregularity of the magnetic field is not quite realistic, because a characteristic scale of galactic inhomogeneities is $\sim 3 \cdot 10^{20}$ cm \cite{kp63}, and is much greater than the critical accretion radius $r_c \sim GM/(u^2 + a_c^2)\,\sim 10^{14}$ cm, when $M \approx M_\odot$, $\sqrt{u^2 + a^2}\sim 10^6$ cm/sec \cite{zn71}. Here $u$ is velocity of a star relative to the gas cloud, and $a_c$ is the sound velocity at the critical point of the flow, where it is equal to the infall velocity of the gas.

 The exact time-dependent solution is obtained for a magnetic field growth for a spherically symmetric accretion  into a BH with a Schwarzschild metric \cite{bkr74}, see also \cite{bk79}. The magnetic field, homogeneous at the initial moment, and ‘frozen’ in the matter, was considered. It was shown in \cite{ginz64,ac70} that the  magnetic field of
a collapsing star produced by currents inside it, dies with time as ($B \simeq t^{-1}$) for a distant observer.
 At a spherical accretion onto the BH, a frozen magnetic field, according to the exact solution, is growing with  time and tends to a quasi-radial structure. The non-stationary behavior of the magnetic field in a stationary flow is connected with infinite conductivity. The magnetic field dissipation
 for chaotic magnetic structures leads to formation of a stationary flow.
  In  section 3 estimations are done for the flux properties in the stationary regime. Equipartition is assumed between
magnetic and kinetic energies of the infalling matter.  The intensity of the magnetic bremsstrahlung radiation of accreting matter, and the spectral energy distribution of the radiation, were roughly determined \cite{bkr74} within this assumption.  The heating associated with annihilation of the magnetic field was taken into account, and the equation of state for a mixture of non-relativistic nuclei and relativistic electrons for
$T\geq 6\cdot 10^{9}$ K was used in addition to suggestions of \cite{shw71}.
It was shown
that the velocity of the infalling matter is less than that of free fall, and the energy output due to the accretion may approach $\sim 0.3\, \dot M\, c^2$. The main contribution to the radiation is given by the essentially relativistic region at $r \approx (2 - 7)r_g$.
The spectrum of radiation of the accretion of magnetized gas into a single BH,  for the chaotic
magnetic field in the gas, is considered in sect. 3.

When BH is in the binary, the falling matter has large angular
momentum and forms a disk around it. The accretion disk
 around a Schwarzschild BH radiates $\sim 0.06\, \dot Mc^2$  and its luminosity reaches $0.42\,\dot Mc^2$ for the limiting Kerr metrics. The magnetic field  may influence
the spectrum of radiation and the physical processes in the accreting disk. It was suggested also that
magneto-rotational instability may be responsible for a generation of the turbulence in the Keplerian accretion disk \cite{bh98}.  An interesting effect was found for a case of
a disk accretion onto a BH in a binary system. In fact, a source with almost black body radiation appears due to gas, falling into BH from the last stable orbit. Its temperature is about $(2 - 3) \cdot 10^7$ K, and luminosity is comparable to that of the whole disk \cite{bkr74}. The presence of rotation diminishes this effect, and it is completely absent in the extremal Kerr metric. This may be considered as a feature, distinguishing a  rotating BH from a non-rotating one.

The magnetic field is generated in the disk, because of thermoelectrical effects.
If the matter would not fall into the hole, values of the generated magnetic
field could be very large $\sim (10^8\,-\,10^{10})$ G. Falling of matter, and large values of
conductivity permit the field to grow only up to several tens of gauss \cite{bkb77}, see also \cite{ck98,bklb02}.

\subsection{Accretion in the presence of a large-scale magnetic field}

The magnetic field can play an essential role in the establishment of the flow of accretion, and in the transformation of the accretion energy into radiation \cite{agu68,shw71}. In the paper \cite{bkr76} a stationary two-dimensional self-similar magnetohydrodynamic solution for the accretion flow was obtained, without assuming a weakness of the magnetic field, the structure and the luminosity of the accretion region were  investigated. The Newtonian gravity was used, and GR effects were taken into account phenomenologically \cite{shw71,shak72,pr72}.

It was assumed that the magnetic field far from a BH is homogeneous and its influence on the flow is negligible. In the plane perpendicular to the direction of the magnetic field, a dense quasi-stationary disk is formed around a BH, which structure strongly depends on  dissipative processes.
Later such model was considered in \cite{nia03}, where it was called as magnetically arrested disk (MAD).
The structure of a disk and the spectrum of the outcoming radiation had been calculated in \cite{bkr76} for
a laminar disk with the Coulomb mechanism of dissipation, and
for	a turbulent disk.
 The density in the radial non-magnetised gas flow, with an adiabatic power $\gamma=4/3$,  increases eight times in the subsonic region, within the interval from infinity to the critical point, where velocity reaches a sound velocity. The magnetic field, frozen in the matter and homogeneous at infinity, increases with time. The approximate law of field increase in the subsonic region, neglecting  density changes, was obtained in \cite{bkr76} using the equation of induction (the ‘freezing’ equation) and the relation $\upsilon\sim r^{-2}$, as

\begin{equation}
\label{eq1a}
B_r=\left(1+\frac{3\dot M}{4\pi\rho_\infty}\frac{t}{r^3}\right)^{2/3}B_0\cos\theta,\quad
B_\theta=-\left(1+\frac{3\dot M}{4\pi\rho_\infty}\frac{t}{r^3}\right)^{1/3}B_0\sin\theta.
\end{equation}
In the supersonic region $\rho \sim r^{-3/2}$, the motion is close to the free-fall, and the magnetic field increases with time as \cite{bkr74}

\begin{equation}
\label{eq1b}
B_r=\left(1+\frac{3}{2}\frac{ct\sqrt{r_g}}{r^{3/2}}\right)^{4/3}B_0\cos\theta,\quad
B_\theta=-\left(1+\frac{3}{2}\frac{ct\sqrt{r_g}}{r^{3/2}}\right)^{1/3}B_0\sin\theta.
\end{equation}
The weak magnetic field at infinity  increases with time, and begins to influence a flow. As a result, in the region where the field attains the value of
$B\sim \upsilon\sqrt{4\pi\rho}$, the two-dimensional essentially non-spherically symmetric stationary flow is formed.  Equipartition between the magnetic and kinetic energies is established in this flow (see Section 4.1). Initially the stationary flow is formed near BH, because the characteristic time of the magnetic field  increase is minimal there. The zone of the stationary flow increases with time, and reaches the boundary of the supersonic flow region. The flow in the subsonic region then ceases to be exactly spherically-
symmetric. It is clear however,  that in the subsonic region the magnetic energy does not exceed the gravitational energy of the matter, and the Alfven velocity $\upsilon_A$ is less than the sound velocity. As a consequence, in the stationary subsonic flow, a mass flux value $\dot M$ is close to the value of $\dot M$ for the non-magnetized gas. The strength of the magnetic field at the edge of the stationary zone depends on the slowly varying values $\dot M$ and critical sonic radius $r_c$, and does not depend on the strength of the field at infinity. In what follows we shall neglect their changes.
The stationary flow zone increases with time. The stationarity conditions  are determined by the hydrodynamic time $t_s\sim r/\upsilon \sim r^{3/2}$, at $\upsilon \sim r^{-1/2}$ in the supersonic region, and $t_s\sim r^{3}$ at $\upsilon\sim r^{-2}$ in the subsonic region.

\begin{figure}[h]
	\centering
	\includegraphics[width=11cm]{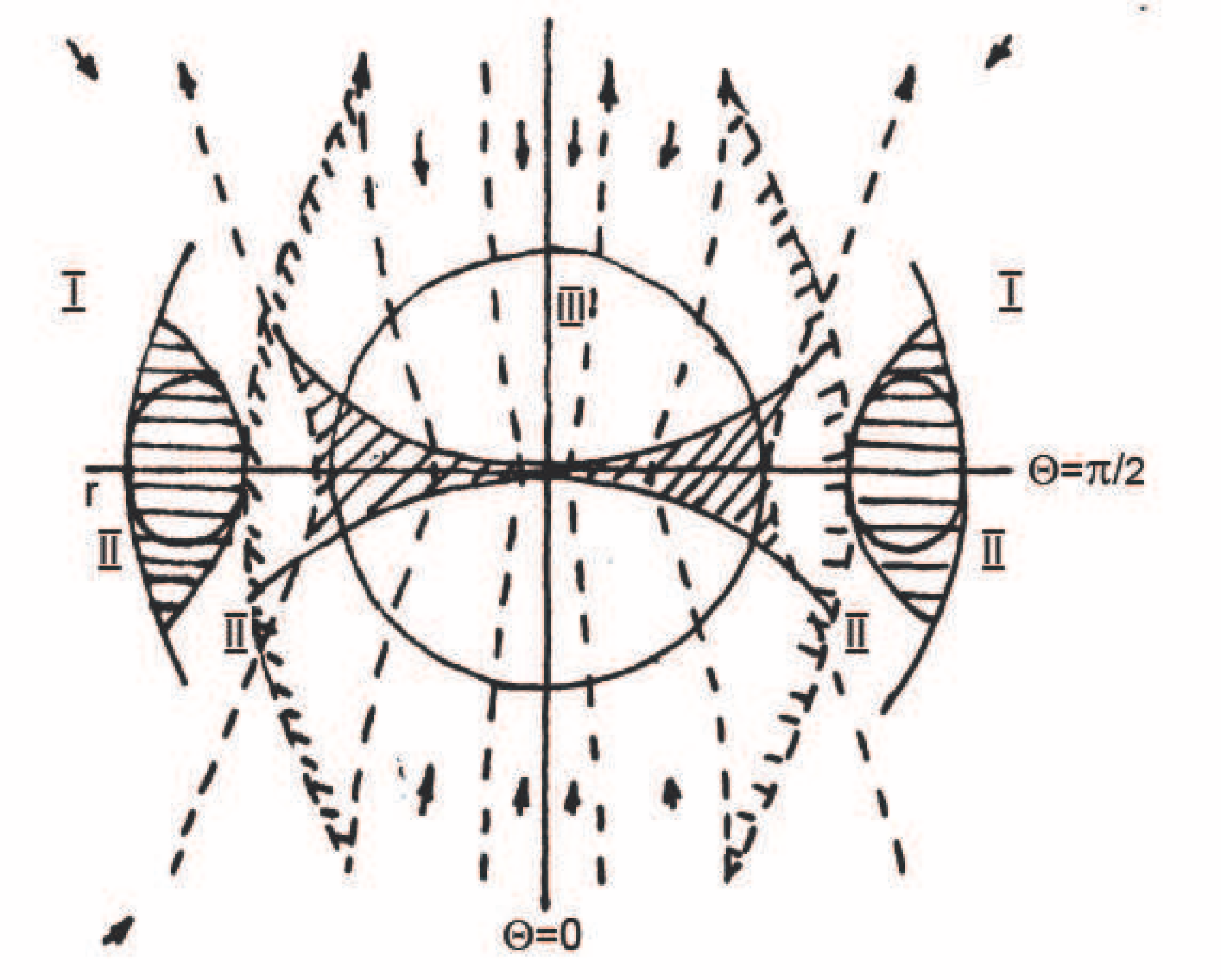}
	\caption{Three zones around a BH during accretion of a gas with a frozen magnetic field, homogeneous at infinity. (I) A zone of
 the stationary hydrodynamic flow with a non-stationary magnetic field. (II) A zone of the stationary flow in which the transition from the subsonic to the supersonic flow takes place. (Ill) A zone of the stationary supersonic flow. Dashed line marks a place where the velocity reaches the critical sound value. At the plane perpendicular to the magnetic field direction at infinity,  thin disk forms around a BH. In zone II the disk becomes thicker and merges with the surrounding flow. Arrows in the direction of the flow velocity have opposite signs in the lower and upper parts, the magnetic field has the same direction there.
  The ring-like stagnation zone formed in the subsonic region of the stationary flow and situated around the symmetry plane is shown by the horizontal strokes.  The sizes of two inner zones are slowly increasing with time, see \cite{bkr76}.
 }
	\label{fig0}
\end{figure}

Finally, the region of the flow can be divided into three zones, see Fig.\ref{fig0}. In a zone around the collapsing star the stationary supersonic flow is established, with a relatively small thermal velocity $\upsilon_T \ll \upsilon\sim \upsilon_A$. The self-similar solution for this zone is obtained in Section 4.2. In the second zone the flow is also stationary, but here a transition from the subsonic flow to the supersonic flow takes place ($\upsilon_T\sim \upsilon_A \simeq \upsilon$). The exact solution in this zone was not obtained analytically. In
the farthest zone the field is growing with time, without influence on the flow pattern, which  is almost spherically-symmetric, see, i.g. \cite{zn71}. In the subsonic region, which is stationary with respect to the magnetic field, the  stagnation zones are formed, in which a matter does not fall into the star and the pressure is balanced by the magnetic field (see Fig.\ref{fig0}). The sizes of two inner zones are slowly increasing with time.
In the section 4.2 a stationary approximate analytic solution for MAD is presented, with application to the  $M_{BH}=10 M_\odot$, representing collapsing stars in the Galaxy.

\noindent
 Phenomena in the accretion  disk supported against gravity by the magnetic field are considered in sect. 4.
In the next two parts results of numerical simulations of MAD are presented, and  magnetohydrodynamic phenomena during accretion of matter with a large-scale magnetic field
onto a magnetized NS are considered, in connection with some particular  X-ray sources.

\section{The non-stationary solution for a magnetic field evolution during a radial accretion onto BH}

\begin{figure}[h]
	\centering
	\includegraphics[width=11cm]{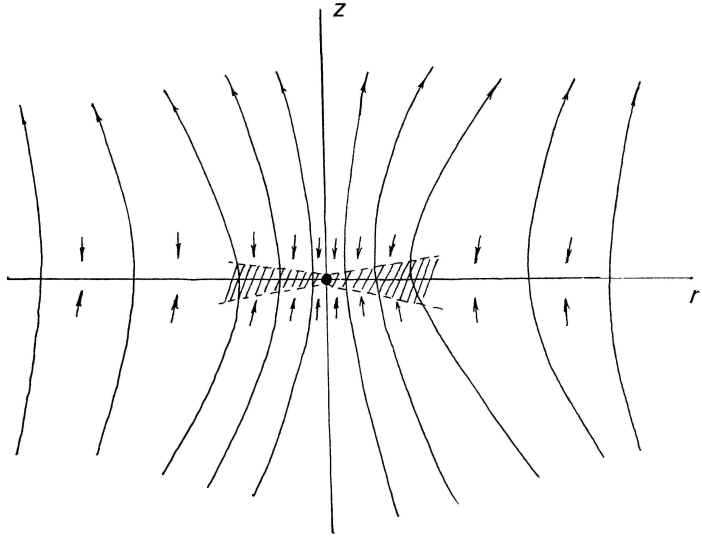}
	\caption{A qualitative  picture of a stationary accretion of matter with a large scale magnetic field onto BH, Arrows indicate the direction of motion of the matter. The magnetic
field far from the star is in the direction of the $z$-axis. The infalling matter forms a disk in the plane $\theta = \pi/2$, which slowly
settles to the star. In the flow region $E_B \sim E_{kin}$, and rotation is entirely absent, from \cite{bkr74}.}
	\label{fig1}
\end{figure}

Let us consider a case, when magnetic field energy density is less than the gravitational one of a gas in the field of a BH, and gas motion is not influenced  by magnetic field. We consider stationary radial gas motion in the BH gravitational field, and find the magnetic field dependence on time using ideal MHD equation \cite{bkr74,bklb02}.
If the 4-velocity of the gas $u^i$, $(i=0,1,2,3)$, moving in the metric $g_{ik}$, $|g_{ik}|=g$ is known, that the equation, determining magnetic field are written as \cite{lich67}

\begin{equation}
\label{eq1}
\frac{\partial}{\partial x^k}\sqrt{-g}(B^i u_k-B^k u^i)=0,
\end{equation}
where $B^i$ is the fourfold magnetic field vector, related to the electromagnetic field tensor $F_{lm}$ by the relation

\begin{equation}
\label{eq2}
B^i=\frac{1}{2\sqrt{-g}}\varepsilon^{iklm} u_k F_{lm}, \qquad B^i u_i=0,
\end{equation}
where $\varepsilon^{iklm}$ is absolutely antisymmetric four-tensor with only 1 and 0 terms.
 The condition of orthogonality $B^i u_i= 0$ indicates that there are only three independent
components of the magnetic field. The relations (\ref{eq1}) and (\ref{eq2}) completely determine time-dependent components of the magnetic field for a given $u^i$, $g_{ik}$ and initial field configuration.
Let us consider the spherically symmetric stationary flow $u^i=(u^0,\, u^r,\, 0,\, 0)$, with initially poloidal magnetic field $B^i= (B^0,\, B^r,\, B^\theta,\, 0)$. In this case Eq. (\ref{eq1}), taking into account the conditions (\ref{eq2}), has a form

\begin{equation}
\label{eq3}
\frac{d}{dt}(\sqrt{-g} u_0^{-1} B^r)=0,
\end{equation}

\begin{equation}
\label{eq4}
\frac{d}{dt}(\sqrt{-g} u_0^{-1} B^\theta)=0,
\end{equation}

\begin{equation}
\label{eq5}
\frac{d}{dt}=\frac{\partial}{\partial t}+\frac{u^r}{u^0}\frac{\partial}{\partial r}.
\end{equation}
Equations (\ref{eq3}) and (\ref{eq4}) are related to the conservation of the magnetic flux along the radial and tangential directions. The solutions of the characteristic system of Eqs. (\ref{eq3})-(\ref{eq5}) is determined by the following integrals

\begin{equation}
\label{eq6}
ct-\int\frac{dr}{u^r/u^0}=c_1\,r_g, \quad \sqrt{-g}u_0^{-1} B^r=c_2\,r_g^2, \quad \sqrt{-g}u^r B^\theta=c_3\,r_g^2.
\end{equation}
The time $t$ enters explicitly in the integrals, so the magnetic field at a given  space point depends on the time, in the stationary hydrodynamic flow. Let us consider Schwarzschild field metric

\begin{equation}
\label{eq7}
ds^2=(1-r_g/r)c^2dt^2-(1-r_g/r)^{-1} dr^2-r^2(d\theta^2+\sin^2\theta\, d\varphi^2).
\end{equation}
The stationary accretion in the Schwarzschild metric, without a magnetic field, has been considered in \cite{mich72}. In the supersonic region $r <r_c$, $r_c$ is the sound velocity radius, the matter velocity may be approximated by a free fall. The components of the four-dimensional velocity are found from the integrals of motion, related to the conservation of energy $E$ and zero angular momentum $L=0$ \cite{ll88}

\begin{equation}
\label{eq8}
E=mc^2 g_{00}\,u^0=mc^2(1-r_g/r)u^0, \quad \sqrt{-g}=r^2\sin\theta.
\end{equation}
For the matter at rest in the infinity, i.e. $E = m\,c^2$, we obtain from Eq.(\ref{eq8}), using $u^i\,u_i = 1$, the relations

\begin{equation}
\label{eq9}
u^0=(1-\frac{r_g}{r})^{-1},\quad u^r=-\sqrt{\frac{r_g}{r}},\quad u_0=1,\quad u_r=\sqrt{\frac{r_g}{r}}(1-\frac{r_g}{r})^{-1}.
\end{equation}
The physical components of the magnetic field ${_{\alpha}} B$ are written as

\begin{equation}
\label{eq10}
_r B=\sqrt{-g_{rr}}\, B^r=(1-r_g/r)^{-1/2}B^r, \quad
_\theta B=\sqrt{-g_{\theta\theta}}\, B^\theta=r\, B^{\theta}.
\end{equation}
Consider the initial field in a form of a uniform magnetic field directed along the $z$ axis. In the Schwarzschild coordinate system the initial field components have a form

\begin{equation}
\label{eq11}
B^r(t=0)=B_0\cos\theta, \quad
r\,B^\theta(t=0)=-B_0\sin\theta\,(1-x), \quad x=\frac{r_g}{r}.
\end{equation}
Performing the integrations in (\ref{eq6}), with account of (\ref{eq9}), and substituting the initial condition (\ref{eq11}) we obtain in the parametric form a solution defining the magnetic field

\begin{eqnarray}
\label{eq12}
\frac{ct}{r_g}+\frac{2}{3}x^{-3/2}+2x^{-1/2}+\ln{\frac{1-\sqrt{x}}{1+\sqrt{x}}}
=\frac{2}{3}x_0^{-3/2}+2x_0^{-1/2}+\ln{\frac{1-\sqrt{x_0}}{1+\sqrt{x_0}}},\nonumber\\
B^r=B_0\cos\theta\frac{x^2}{x_0^2}, \quad
B^\theta=-B_0\sin\theta\,\frac{x^{3/2}}{r_g\,x_0^{1/2}}(1-x_0),
\end{eqnarray}
where $ x_0=\frac{r_g}{r_0}$. To find the physical meaning of the quantity $x_0$ we transfer to the comoving  coordinates system $(\tau,\,\varrho,\,\theta,\,\phi)$, in which the Schwarzschild metric is written as \cite{ll88}

\begin{equation}
\label{eq13}
ds^2=
c^2 d\tau^2-\frac{r_g}{r}
d\varrho^2-r^2(d\theta^2+\sin^2{\theta}\, d\phi^2),
\end{equation}
with
no pathology on the black hole horizon.
The connection between Schwarzschild and comoving coordinates
$(\tau, \varrho)$ (angle coordinates $\theta$, $\phi$ are the same)
is

\begin{equation}
\label{eq14}
c\tau=ct+r_g\biggl[2x^{-1/2}+\ln{\frac{1-\sqrt{x}}{1+\sqrt{x}}}\biggr],
\end{equation}
$$\varrho=ct+ r_g\biggl[\frac{2}{3}x^{-3/2}+2x^{-1/2}+
\ln{\frac{1-\sqrt{x}}{1+\sqrt{x}}}\biggr]
$$
After comparison of (\ref{eq12}) and (\ref{eq14}) we see that the parameter $x_0$ is a function of the Lagrangian coordinate $\varrho$, and, therefore, is itself a  non-dimensional Lagrangian coordinate,  while $r_0$ is a dimensional one. For any given time $t$ the matter cannot reach the horizon $(x=1)$, therefore the magnetic field remains equal to zero there \cite{bkr74}.
In general $_\varrho B$ grows more rapidly than $_\theta B$ i.e., the asymptotic magnetic field becomes close to the radial one. It is easy to show in the simple Newtonian case, when the equations of the lines of force $dr/B^r= r d\theta/B_\theta$ may be easily integrated, giving

\begin{equation}
\label{eq15}
\left(1+\frac{3}{2}\frac{ct\sqrt{r_g}}{r^{3/2}}\right)^{2/3}r\sin\theta={\rm const}.
\end{equation}
At $r\rightarrow\infty$ we obtain the picture of a uniform field $r \sin\theta=$const. For small $r$ (or as $t\rightarrow\infty$) we obtain $\sin\theta\cdot t^{2/3} = {\rm const}$, i.e. a radial magnetic field varying (increasing) with time. In Newtonian limit we obtain the following solution for the  evolution with time of the physical components of the magnetic field

\begin{eqnarray}
\label{eq16}
x_0=x\,\left(1+\frac{3}{2}\frac{ct\sqrt{r_g}}{r^{3/2}}\right)^{-2/3},\quad
_\varrho B=\left(1+\frac{3}{2}\frac{ct\sqrt{r_g}}{r^{3/2}}\right)^{4/3}B_0\cos\theta,\\
_\theta B=-\left(1+\frac{3}{2}\frac{ct\sqrt{r_g}}{r^{3/2}}\right)^{1/3}B_0\sin\theta.
\end{eqnarray}
The solution given by Eqs. (\ref{eq12}),(\ref{eq16})  was obtained for a given radial flow of matter without account of the magnetic field influence on the motion. Actually, the action of the increasing  magnetic field terminates its growth, and the solution approaches to the stationary self-consistent regime.
Increase of a magnetic field energy, mainly of the radial component, has a kinetic energy as a source. Therefore in the self-consistent quasi-stationary regime we expect the equipartition between these two energies. For a low-temperature gas the radial flow velocity $\upsilon_r$ is proportional to the free-fall one, and the magnetic energy density is equal to the kinetic energy one with gas density $\rho$. Finally, we have the following estimation for the maximum of the radial magnetic field in the flow with a mass flux $\dot M$ to BH with a mass $M$ as \cite{bkr74}, see numerical estimations in Sect. 4.3,

\begin{equation}
\upsilon_r =\alpha c\sqrt{\frac{r_g}{r}},\,\,\, \rho=\frac{1}{\alpha}\frac{\dot M}{4\pi c r^2}\sqrt{\frac{r}{r_g}},\,\,\,(B_r^2)_{max} = \rho \upsilon_r^2=\alpha\frac{\dot M c}{r^2}\sqrt{\frac{r_g}{r}}. 	\label{refs3}
\end{equation}
We consider that the lines of force are not gathered in to the centre, but are distributed along the disk, forming in the plane $\theta = \pi/2$, and expanding outward. An approximate diagram of the lines of force is given in Fig. \ref{fig1} from  \cite{bkr74}. More detailed considerations of the established flow due to accretion in the presence of a magnetic field is given in the section 4.
Let us estimate the quantities characterising the accretion with a chaotic magnetic field, and the magnetic bremsstrahlung radiation during such accretion onto a BH.

\section{Radiation flux and spectrum at a radial accretion of matter with chaotic magnetic field onto a black hole }

In the case of a spherically symmetric accretion, the flow passes through
the sound-velocity point at a saddle-like singularity of the system of
hydrodynamical equations \cite{306b}. The condition for the flow to pass through
the sound-velocity point uniquely  determines the mass flux $\dot M$
and all the flow properties for given $T_\infty$ and $\rho_\infty$. An
adiabatic flow in the point-mass gravitational field can pass through
the sound-velocity point only if the adiabatic index $\gamma<5/3$.
The theory of adiabatic accretion is given in \cite{306b,227}, and is similar
to the stellar wind theory. We consider here the
flows passing through the sound-velocity point and supersonic near the centre
of gravity.
 In presence of a magnetic field more singular points may appear in the accretion flow, connected with magneto-sonic and Alfven velocities, see \cite{gam99,mm18}. In our consideration the transition through the gas sonic point happens far from BH and equipartition zone, where magnetic field influence may be neglected.

Studies of spherically symmetric accretion of interstellar gas onto
black holes have shown that in the presence of a magnetic field frozen in
plasma, the efficiency of the kinetic energy conversion into heat approaches
$\eta\approx 10\%$ \cite{shw71}, while in the absence of a field the bremsstrahlung
efficiency is $\eta \sim 10^{-8}$. When the gas flows radially, the lines of
magnetic force stretch along a radius, $B_r\sim r^{-2}$, and the magnetic
energy per unit volume $E_M\sim B^2\sim r^{-4}$ increases more rapidly than
the kinetic energy $E_{\rm kin}\sim\rho \upsilon^2\sim \dot M\upsilon/r^2\sim r^{-5/2}$
($\dot M=4\pi\rho \upsilon r^2$ is the stationary mass flux, free-fall velocity
$\upsilon\sim r^{-1/2}$). Since the energy $E_M$ cannot physically exceed
$E_{\rm kin}$, it is assumed in \cite{shw71} that an equipartition of energy
$E_M\approx E_{\rm kin}$ is supported by the dissipation of magnetic energy,
the excess of which is consumed by plasma heating. This heating was taken
into account in \cite{bkr74} and leads to an increase in efficiency $\eta$ to
$30\%$ that may be considered as a realistic estimate under these assumptions.

If $E_M\sim r^{-4}$ is the magnetic field energy  with no dissipation,
and $E'_M=E_{\rm kin}\sim r^{-5/2}$ is the energy of the magnetic field in a flow,
then an increase of entropy per unit volume along a radius due to the field
annihilation in a stationary flow is given by
\begin{equation}
  {Q_M}={\left(\rho T\,\frac{dS}{dr}\right)_M=
  \left(\frac{dE_M}{dr}-\frac{dE'_M}{dr}\right)_{E_M=E'_M}}
  ={-\,4\,\frac{E_M}{r}+\frac{5}{2}\,\frac{E_M}{r}=-\frac{3}{2r}\,\frac{B^2}{8\pi}}\,.
\label{eq:11.3.1}
\end{equation}

\noindent Consider separately the regions with non-relativistic
electrons with
\[
  kT\ll\mec{}2,\quad \gamma_1=5/3,
\]

\noindent and relativistic electrons with\footnote{\noindent Protons
are always non-relativistic.}
\begin{equation}
  {kT}\gg {\mec{}2}\,,\quad
{\rho E}=\frac{1}{2}\left(3P+\frac{3}{2}\,P\right)=\frac{9}{4}\,P=nP\,,\quad
{\gamma_1}=1+\frac{1}{n}=\frac{13}{9}\,.
\label{eq:11.3.2}
\end{equation}

\noindent Here, $P\r e=P\r p=P/2$, $\gamma_1$ is the adiabatic power,
 $n$ is the adiabatic index, for simplicity
we consider hydrogen plasma. From the energy balance equation
\begin{equation}
  \frac{dE}{dr}-\frac{P}{\rho^2}\,\frac{d\rho}{dr}=\frac{Q_M}{\rho}
  -\frac{\varepsilon_B}{\upsilon_r}\,,
\label{eq:11.3.3}
\end{equation}

\noindent where $\varepsilon_B$ (${\rm erg~g^{-1}~s^{-1}}$) is a rate of
magneto-bremsstrahlung losses of the Maxwell plasma, with
\begin{equation}
  \frac{B^2}{8\pi}=\frac{1}{ 2}\,\rho \upsilon_r^2,\quad
  \upsilon_r=\alpha \upsilon_{ff}=\alpha\sqrt{\frac{2GM}{ r}},\quad
  \overline{B_\perp^2}=\frac{2}{ 3}\,\overline{B^2}\,,
\label{eq:11.3.4}
\end{equation}

\noindent
and \cite{bkf69,bkr74}

\begin{equation}
\label{ebnr}
\varepsilon_B=2
\frac{\ee^2}{m\r pc}
  \left(\frac{eB_\perp}{m_e c^2{}{}}\right)^2
  \frac{kT}{m_e c^2}
  \approx 0.46\,TB_\perp^2~{\rm erg~g^{-1}~s^{-1}} \quad
  {\rm for~}\quad kT\ll m_e c^2 \quad {(\rm NR)},
\end{equation}
\begin{equation}
\label{ebr}
\varepsilon_B=8\,\frac{\ee^2}{m\r pc}
  \left(\frac{\ee B_\perp}{m_e c^2}\right)^2
  \left(\frac{kT}{m_e c^2}\right)^2
  \approx 3.2\times 10^{-10}\,T^2B_\perp^2\,\,{\rm erg\,g^{-1}\,s^{-1}}
  \quad {\rm for}\quad kT\gg m_e c^2 \quad {(\rm R)}.
\end{equation}

\noindent
We obtain equations for $T(r)$ in the form
\begin{equation}
  \frac{3}{2}\,\frac{dT}{dr}+\frac{3}{2}\,\frac{T}{r}+
  \frac{3}{4}\,\frac{\alpha^22GM}{{\cal R}_gr^2}-
  1.5\,\frac{T\dot M}{{\cal R}_gr^2}=0
  \quad{\rm (NR)}
\label{eq:11.3.7}
\end{equation}
  \begin{equation}
  \frac{9}{4}\,\frac{dT}{dr}+\frac{3}{2}\,\frac{T}{r}+
  \frac{3}{4}\,\frac{\alpha^22GM}{{\cal R}_gr^2}-2.2\times 10^{-10}\,
  \frac{T^2\dot M}{{\cal R}_gr^2}=0. \quad{\rm (R)}
\label{eq:11.3.8}
\end{equation}

\noindent Here ${\cal R}_g=2k/m_p$ is the gas constant for ionized hydrogen.
For given values of $\rho_\infty$, $T_\infty$, $M$, the mass flux is
determined by

\begin{equation}
  \dot M=4\pi\rho \upsilon_rr^2=\frac{10^{32}}{c^2}\MM2\rhoinf
 \times\TK{-3/2}~{\rm g~s^{-1}}\,.
\label{eq:11.3.9}
\end{equation}

\noindent Neglecting radiation in Eq.~(\ref{eq:11.3.7}) and adiabatic heating in Eq.~(\ref{eq:11.3.8}),
we obtain the solution in the form
\begin{equation}
  T=2\times 10^{12}x+2.7\times 10^{12}x\alpha^2
  \ln\left(\frac{10^8x}{T_\infty/10^4~{\rm K}}\right)
  \quad {\rm (NR)}
\label{eq:11.3.10}
\end{equation}
\begin{equation}
  T=\frac{m_e c^2{}2}{k}+T_1\frac{e^{a(x-x_0)}-1}{e^{a(x-x_0)}+1},\quad
  T_1\gg \frac{m_e c^2{}2}{k}
  \quad {\rm (R)}\,.
\label{eq:11.3.11}
\end{equation}

\noindent Here,
\begin{equation}
\begin{array}{lcl}
  \displaystyle{x}&=&\displaystyle{r_g/r=\frac{2GM}{rc^2}<1}\,,
  \\[12pt]
  \displaystyle{T_1}&=&\displaystyle{2.8\times 10^{12}\alpha\left(\frac{T_\infty}{10^4}\right)^{3/4}
  (M/M_\odot)^{-1/2}\left(\frac{\rho_\infty}{10^{-24}}\right)^{-1/2}}\,,
  \\[12pt]
  \displaystyle{a}&=&\displaystyle{1.3\,\alpha\MM{1/2}\TK{-3/4}\rhoinf}\,.
\end{array}
\label{eq:11.3.12}
\end{equation}

\noindent A value of $x_0$ slightly depends  on $T_\infty$, $\rho_\infty$,
and for various values of $\alpha$ it is shown in Table\ref{tab1}.
 When $x=x_0$, we have $T=m_e c^2{}2/k$, and the solutions (\ref{eq:11.3.10}) and
(\ref{eq:11.3.11}) fit to one another. The luminosity and spectrum for this
model were calculated in \cite{bkr74}. The luminosity due to
magneto-bremsstrahlung is determined mainly by relativistic electrons from
(\ref{eq:11.3.11}), giving

\begin{table}
\caption{Dependence of $x_0$ on $\alpha$, from \cite{bkr74}}

\medskip

\centerline{
\begin{tabular}{c|c|c|c}
$\alpha^2$  &1 &1/3 &1/10\\
\hline
& & & \\[-6pt]
$x_0$  &$2\times 10^{-4}$ &$5\times 10^{-4}$ &$1.2\times 10^{-3}$
\end{tabular}}
\label{tab1}
\end{table}

\medskip

$$
 L_B=2.7\times 10^{31}\alpha^4(M/M_\odot)^3\,(\rho_\infty/10^{-24}\,{\rm g\,cm}^{-3})
 (T_\infty/10^4{\rm K})^{-3}, \quad a<1
$$
\begin{equation}
 L_B = 9\times
10^{31}\alpha^2(M/M_\odot)^2{(\rho_\infty/10^{-24}~{\rm
g~cm}^{-3}})
(T_\infty/10^4{\rm K})^{-3/2}, \quad a\gg 1
\end{equation}

\begin{figure}
\centerline{\includegraphics[width=11cm]{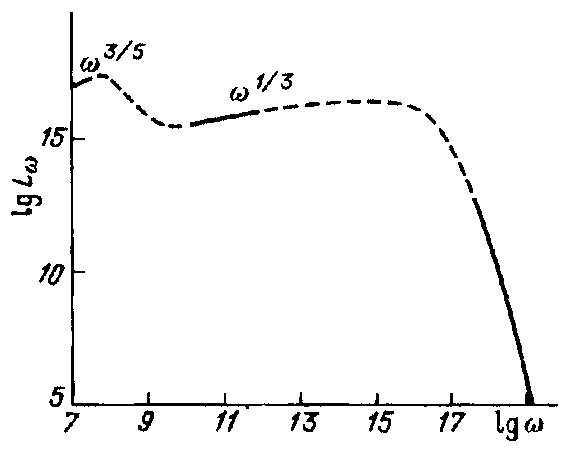}}
\caption{Magneto-bremsstrahlung spectrum of a BH of $M=10\,
M_\odot$ for a spherically symmetric accretion and random magnetic
field at $\rho_\infty = 10^{-24}$ g\,$\cdot$\,cm$^{-3}$,
$T_\infty = 10^4$~K, $\alpha^2 = 1/3$. The  solid lines
represent asymptotic dependencies, dashed lines give
extrapolations, from \cite{bk11}.}
\label{fig2}
\end{figure}

\noindent
From comparison of this expression with (\ref{eq:11.3.9}), we find that for a
realistic value $\alpha^2=1/3$ the quantity $\eta=L_B/\dot Mc^2\le
30\%$. An approximate BH emission spectrum $L_\omega$
($L_B=\int_0^\infty L_\omega\,d\omega$) for mass
$M_{BH}=10 M_\odot$ is given in Fig.~\ref{fig2} from \cite{bk11}. The range
with $L_\omega \sim \omega^{3/5}$ is related to the emission of
non-relativistic electrons; at $kT\gg m_e c^2{}2$, $\hbar\omega\ll kT$,
$L_\omega\sim\omega^{1/3}$, while at $\hbar\omega\approx
\hbar\omega_{B,{\rm max}}(kT_1/m_e c^2{}2)^2\sim 10~{\rm eV}$ for
$\rho_\infty=10^{-24}~{\rm g~cm^{-3}}$, $T_\infty=10^4~{\rm K}$,
$B\r{max}\sim 10^5~{\rm G}$ an exponential cut-off
occurs.

It was obtained in \cite{bkr74}, that the exponential cutoff frequency of brems\-strahlung radiation  $\omega_{cut}$, of thermally heated gas, has the following dependence on accretion parameters

\begin{equation}
\omega_{cut}\sim \alpha^{5/2} T_\infty^{3/4} \rho_\infty^{-1/2} M^{-1},
\label{refs32}
\end{equation}
so for the same surrounding  gas the cut-off frequency is inversely proportional to a BH  mass. We may conclude therefore, that the observed high energy radiation from a super-massive BH in AGNs is formed due to strong non-thermal heating of electrons around BH or/and in jets, observed in these objects.

 The synchrotron
radiation spectrum of a unit volume of relativistic Maxwell
plasma is given by \cite{bkf69}
\[
  I_\omega=
    \frac{\sqrt2 \rho e^2}{6 \mu_Z m\r p c}\,
    \omega_B z\exp\left[-\left(\frac{9}{2}\right)^{1/3}z^{1/3}\right],
    \quad z\gg 1
   \]
   \begin{equation}
    ={\frac{3^{1/6}}{\pi}}\Gamma(4/3)\Gamma(5/3)\,{\frac{\rho e^2}{\mu_Z m\r p c}}\,
    \omega_B z^{1/3},\quad z\ll 1
\label{ssin}
   \end{equation}
  \[
  z=\frac{\omega}{\omega_B}\left(\frac{m_e c^2}{kT}\right)^2,\quad
  I~({\rm erg~cm^{-3}~s^{-1}})=\int_0^\infty I_\omega\,d\omega.
  \]
The visible magnitude $m_\upsilon$ for such a BH is given by \cite{bkr74}
\[
  m_V=4.8-2.5\log L/\Lo+5\ln(R/10~{\rm pc})
  \approx 14.1-7.5\log M/M_\odot
  \]
\begin{equation}
  -2.5\log\left[\rhoinf^{3/2}\TK{-9/4}\right]
  +5\log\left(\frac{R}{10~{\rm pc}}\right)\,.
\label{eq:11.3.14}
\end{equation}

\noindent At increasing of luminosity, connected with an increase of a gas density and (or) BH mass, the interaction of the outgoing bremsstrahlung radiation flux with an accretion flow is becoming essential.
Calculations of accretion into a BH including the reciprocal effect
of radiation had been done in \cite{par90,park90}. For a case of accretion onto a NS, a
similar effect had been investigated in \cite{bkb80}.

\section{Accretion at an ordered magnetic field}

If the characteristic scale of non-uniformity of magnetic field
is much larger, that the accretion radius
\begin{equation}
  r\r a=\frac{GM}{\upsilon_s^2},\quad
  \hbox{where $\upsilon_s$ is the sound velocity in the gas}\,,
\label{eq:11.3.15}
\end{equation}

\begin{figure}
\centerline{\includegraphics[width=11cm]{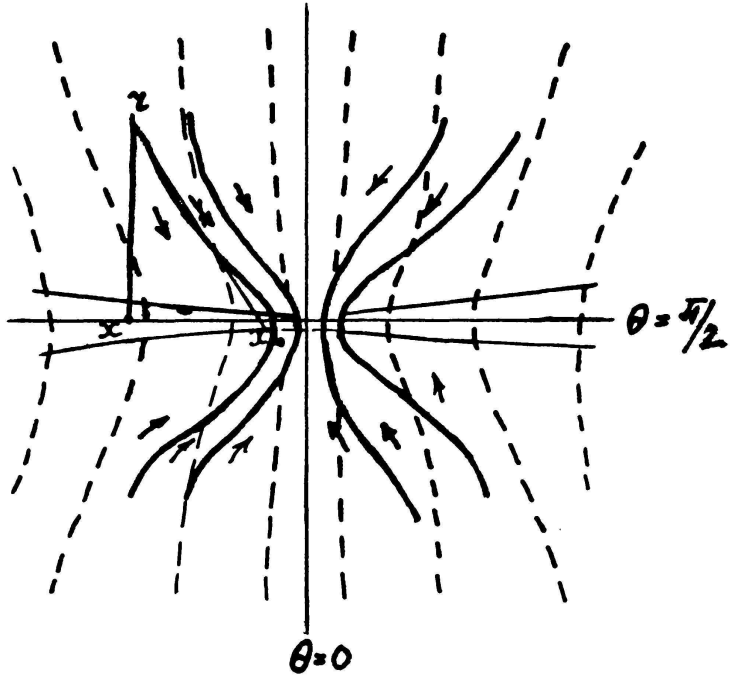}}
\caption{Schematic pattern of magnetic field lines in the matter
around a BH for a field which is uniform at infinity, with the
inclusion of distortions due to disk currents. The non-perturbed external magnetic field is
shown by dashed lines. The solid lines determine the magnetic field lines, influenced by the
azimuthal electrical currents in the disk. The domain of the disk influence is determined by the
radius $r(x_0)$, $x(x_0)$, where field perturbations from the disk are small.
The {\sl arrows}
indicate a direction of a gas flux velocity, with account of perturbations from the disk,
from~\cite{bkr76}.}
\label{fig3}
\end{figure}

\noindent the flow looses its spherically symmetry. For
a uniform magnetic field the accretion symmetry is cylindrical. If
a BH is at rest, a stationary pattern of magnetic lines is established,
the gas flows along them  and forms a disk in the plane of symmetry.
A qualitative picture of the flow is shown in Fig.~\ref{fig3}. At a finite
conductivity in the disk a gas infiltrates through force lines of magnetic
field towards a BH. The formation process,
the structure of a disk supported by a magnetic field and its radiation
had been considered in \cite{bk79,bkr76}.

\subsection{Self-similar solution for the stationary flow outside the symmetry plane }

Let us consider a flow in the inner supersonic region, where pressure is negligible and such a flow is essentially directed by the magnetic field. In the axial symmetry case, the stationary picture will be two-dimensional. The basic equations for two-dimensional accretion onto a gravitating centre (BH) with mass $M$, of the magnetized gas with perfect conductivity, and without pressure are written in the form \cite{bkr76}

\noindent
\begin{equation}
 \upsilon_r\, \p{\upsilon_r}r+\frac{\upsilon_\theta}{r}\, \p{\upsilon_r}\theta-\frac{\upsilon_\theta^2}{ r}=-\frac{GM}{r^2}
  -\frac{B_\theta}{4\pi\rho r}\left[\frac{\partial(rB_\theta)}{\partial r}-\frac{\partial B_r}{\partial \theta}\right],
\label{bac1}
\end{equation}

\begin{equation}
 \upsilon_r\, \p{\upsilon_\theta}r+\frac{\upsilon_\theta}{r}\, \p{\upsilon_\theta}\theta+
 \frac{\upsilon_r \upsilon_\theta}{r}=
\frac{B_r}{4\pi\rho r}\left[\frac{\partial(rB_\theta)}{\partial r}-\frac{\partial B_r}{\partial \theta}\right],
 \label{bac2}
\end{equation}

\begin{equation}
 \frac{1}{r} \p{}r(r^2 \rho \upsilon_r)+\, \frac{1}{\sin\theta}\p{}\theta (\sin\theta\rho \upsilon_\theta)
 =0\,,
\label{bac3}
\end{equation}

\begin{equation}
 \frac{1}{r}\, \frac{\partial}{\partial r}(r^2 B_r)+\frac{1}{\sin\theta}\p{}\theta(\sin\theta\, B_\theta)=0\,,
\label{bac4}
\end{equation}

 \begin{equation}
\upsilon_rB_\theta-\upsilon_\theta B_r=0.
\label{bac5}
\end{equation}
We assumed that a picture of the flow is stationary $(d/dt = 0)$, two-dimensional $\partial/\partial\phi = 0$, and the matter falling to a BH has no angular momentum $\upsilon_\phi = B_\phi = 0$.
Eqs. (\ref{bac1})-(\ref{bac2}) are the $r$, and $\theta$ components of the Euler equation; (\ref{bac3}) is the  continuity equation; (\ref{bac4}) defines zero divergency of the magnetic field (${\rm div} {\bf B}=0$), and (\ref{bac5}) defines the ‘freezing’ of the magnetic field, when the flow patterns are parallel to the magnetic field lines. The magnetic force on the right sides of Eqs. (\ref{bac1}),(\ref{bac2}) is perpendicular to the streamlines. The Bernoulli equation is obtained by multiplying (\ref{bac1}) by $\upsilon_r$, (\ref{bac2}) by $\upsilon_\theta$ and combining them. It does not contain the magnetic field: i.e.,

\begin{equation}
\upsilon_r \frac{\partial\upsilon^2}{\partial r} + \frac{\upsilon_\theta}{r}
  \frac{\partial\upsilon^2}{\partial\theta}= - 2\upsilon_r \frac{GM}{r^2}
\label{bac6}
\end{equation}
We did not find a general solution of the system (\ref{bac1})-(\ref{bac5}), but have found a self-similar solution of this system in the form

\begin{equation}
\upsilon_r =	-\sqrt{2GM/r} \, f(\theta),	\quad
\upsilon_\theta = \sqrt{2GM/r} \, g(\theta), \quad \rho=\rho(\theta),\quad
{\bf B}=a\rho{\bf v},\,\,\,	a ={\rm	const}.	
\label{bac7}
\end{equation}
The last relation (\ref{bac7}) is connected with a coincidence of
the magnetic field and the streamlines directions, follows from comparison of Eqs. (\ref{bac3}) and (\ref{bac4}). Substituting two first relations from (\ref{bac7}) into Eq.(\ref{bac6}) we find that it is satisfied when

\begin{equation}
f^2+g^2 =	1.	
\label{bac8}
\end{equation}
 It is convenient, using (\ref{bac7}),(\ref{bac8}), to transfer from ($f,g,B,\upsilon$) to new variables ($z,y$)

\begin{equation}
f = \cos z,\quad g = \sin z\quad y=\frac{a^2}{4\pi}\rho=\frac{B^2}{4\pi\rho\upsilon^2}.
\label{bac9}
\end{equation}
 Here $y$  is equal to the ratio of the magnetic and kinetic energies. For $y$ and $z$, we have two equations of the first order

\begin{equation}
\frac{dy}{d\theta}=y(1-y)\frac{\cot z - \cot \theta}{\sin^2z-y}\cdot \sin^2{z},
\label{bac10}
\end{equation}

\begin{equation}
\frac{dz}{d\theta}=\frac{1}{2}-
\frac{\sin z \cos z (\cot z - \cot\theta)}
	{\sin z - y} y
\label{bac11}
\end{equation}
 A solution for zero magnetic field is $y = 0$, $z = \theta/2$. Using Eq. (\ref{bac3}), instead of the third relation (\ref{bac7}), we obtain the solution in the form \cite{bkr76}

\begin{equation}
\upsilon_r = - \sqrt{2GM/r}\cos(\theta/2), \quad
\upsilon_\theta = \sqrt{2GM/r} \sin(\theta/2), \quad
\rho(r,\theta)=\rho_0 \Phi\left(\frac{r}{R}\sin^2\frac{\theta}{2}\right)\tan{\theta/2},
\label{bac12}
\end{equation}
 where $\Phi$ is an arbitrary function.
This solution describes axially symmetric streamline around a gravitational center of particles flow having an angular momentum at infinity, see, for instance, \cite{spi}.
For an example, at $\Phi(x)=1/x$ we have the density distribution $\rho(r,\theta)= \frac{\rho_0}{cos(\theta/2)}\sqrt{\frac{R}{r}}$, describing the flow of a cold gas, with uniform density at infinity, into rapidly moving  gravitational center. A self-similar solution for such a flow, with a shock wave behind the center of gravity, was obtained in \cite{bkkk79}.
We are interested, however, in a solution with the spherically-symmetric distribution of matter at infinity. Deviations from the spherical symmetry around a BH appear only due to the magnetic field.
In a general case the numerical solution of the system (\ref{bac10}),(\ref{bac11}) is obtained using an expansion near the singular points of the system. This system have three singular points corresponding to the simultaneous zero's of the numerators and denominators in the right-hand sides of Eqs. (\ref{bac10}),(\ref{bac11})

\begin{equation}
{\bf 1)}\,y_c=0,\,\,z_c=k\pi;\quad {\bf 2)}\, y_c=1,\,\,z_c=\pi/2+k\pi;
\quad {\bf 3)}\, y_c=\sin^2 z, \,\,z_c=\theta+k\pi,\,\,k-0,1,2...
\label{bac13}
\end{equation}
The first two points are special cases of the third one, when $\theta = 0,\, \pi/2$, but in a view of simplicity and physical clarity they are listed separately. The case $y_c=0$, $z_c=k\pi$ corresponds to a non-magnetized radial flow of  cold gas, with the solution $y=0,\,\,g=0,\,\,f=1,\,\,\upsilon_\theta=0,\,\,\upsilon_r=-\sqrt{2GM/r}$, describing a free-fall velocity flow from infinity.
  An equation for the singular point at $\sin^2 z_c - y_c = 0$ corresponds to the relation between physical quantities as

 \begin{equation}
\upsilon_\theta = \pm \frac{B}{\sqrt{4\pi\rho}},	
\label{bac14}
\end{equation}
what determines an equality between $\upsilon_\theta$ and the Alfven velocity in the singular point. Physically relevant solution is described only by real functions. It was shown in \cite{bkr76}, using expansion of the solution in this singular point, that real coefficients of expansion exist only for $y_c=1$, and have a complex character at $y_c<1$. Therefore the only physical solution of Eqs. (\ref{bac10}),(\ref{bac11}) goes through the singular point where $\upsilon_r=0$, and Alfven velocity is equal to the full gas physical velocity.
 An absence of stationary solutions at $y=\frac{a^2}{4\pi}\rho=\frac{B^2}{4\pi\rho\upsilon^2}<1$ is related to amplification of weak fields in non-stationary regime. It is worth noting, that in motion of a cold gas a relation between variables in the singular point are valid for the whole flow, therefore
we consider the solution with $y \equiv 1$,  i.e. an equipartition between magnetic and kinetic energies everywhere. In this case only one equation remains

\medskip
\begin{table}
\caption{Self-similar solution for $y=1$, from \cite{bkr76}}
   \begin{tabular}{c|c|c|c|c|c|c|c|c|c}
$\theta$\,\,&$10^{-3}$ &0.2\,\,\,&0/4\,\,\,&0.6\,\,\,&0.8\,\,\,&1\,\,\,\,&1.2\,\,\,&1.4\,\,\,&$\pi/2\,\,$\\
\hline
& & & & & & & & & \\[-6pt]
$z$\,\,&$7.5\cdot 10^{-4}$ &0.15\,\,&0.30\,\,&0.46\,\,&0.62\,\,&0.78\,\,&0.96\,\,&1.2\,\,\,&1.4\,\,
    \end{tabular}
    \label{tab2}
\end{table}

\begin{equation}
\frac{dz}{d\theta}= \frac{3}{2} - \tan z \cdot \cot \theta;\quad {\rm with\,\,the\,\,boundary\,\,
 condition} \quad z(0)=0,
\label{bac15}
\end{equation}
corresponding to the radial motion of the matter on the pole. The numerical
solution of Eq. (\ref{bac15}) in the region $0\le<\theta\le\pi/2$ is given in Table \ref{tab2}. The density in this solution is constant $\rho=\rho_0$, and the velocity $\upsilon$ is equal to the Alfven velocity
$\upsilon_A=\frac{B}{\sqrt{4\pi\rho}}$.

For small $\theta$  the following expansion goes from Eq.~(\ref{bac15})
$z = 3/4\theta + ...$. The pattern of the streamlines, coinciding with  magnetic force lines is shown schematically  by dashed lines in Fig.~\ref{fig3}, see \cite{bkr76}.
Equation (\ref{bac15}) does not change under a mirror transformation $\theta\rightarrow \pi-\theta$ and $z\rightarrow\, - z$. So, the solution is anti-symmetrically extended to the lower hemisphere $\pi/2< \theta\le \pi$, with $\upsilon_r$ having the same sign as at the upper hemisphere (directed to the centre,) and $\upsilon_\theta$  changes its sign (see Fig.\ref{fig3}).  Therefore the plane $\theta=\pi/2$ occurs to be  singular.  In this plane, a quasi-stationary disk is formed, in which matter moves to a BH, penetrating through magnetic field lined due to finite electrical conductivity.

\subsection{Stationary accretion disk in presence of a large-scale magnetic field}

We study a structure of a quasi-stationary magnetized accretion disk using a simplified approach \cite{bkr76}.
To obtain an analytic solution for the accretion disk structure supported by magnetic field is possible only when we neglect centrifugal forces from rotation. Account of both forces in construction of the model is possible only by numerical simulations. Some results of simulations of this model are presented is Sect.5, where more references are given. It follows from simulations presented in \cite{ig08}, that in the presence of the vertical magnetic field angular momentum is effectively extracted from the falling flow with a Keplerian rotation on the outer boundary, so that in the inner region around BH magnetic forces exceed the centrifugal barrier. The solution shown below is valid therefore for the inner region around a BH. The extended transition zone between the outer Keplerian accretion disk at $r \ge (100-1000)\, R_g$, and inner, magnetic force dominated region at $r\le (20-40)\, R_g$ may be studied quantitatively only by numerical simulations.

Let us consider an equilibrium of a non-rotating disk with a balance between magnetic forces and
gravity:

\begin{equation}
  \frac{GM\Sigma}{r^2}=\frac{1}{c}\,B_\theta I_\varphi\approx
  \frac{2\pi}{c^2}\,I_\varphi^2\,.
\label{eq:11.3.16}
\end{equation}

\noindent Here, $\Sigma=2h\rho$ is the surface density, $\rho$ is the average
density of the gas, $I_\varphi$ is a circular electrical current surface density.  We
have roughly \cite{bkb72}
\begin{equation}
  B_\theta\approx B_r\approx \frac{2\pi}{c}\,I_\varphi\,.
\label{eq:11.3.17}
\end{equation}

\noindent Equilibrium along $z$ axis is supported by a balance between the vertical pressure gradient, and gravity
\begin{equation}
  \frac{dP}{dz}=-\frac{\rho GM}{r^2}\,\frac{z}{r},\quad
  h\approx\left(\frac{r^3}{GM}\,\frac{P}{\rho}\right)^{1/2}\,.
\label{eq:11.3.18}
\end{equation}

\noindent The disk heating  due to extraction of the gravitational energy, is related to a slow motion of the disk into a BH, and due to additional heating by matter falling to the disk along magnetic field lines, at an almost free-fall velocity.
Finally we obtain an expression for the energy flux from unit disk surface in the form

\begin{equation}
  F=\frac{GM\dot M}{4\pi r^3}\,
  \left[1+\frac{1}{2}\left(\frac{r}{R}\right)^{3/2}\right]\,.
\label{eq:11.3.19}
\end{equation}

\noindent From the mass conservation law we obtain an expression for the radial velocity of the disk matter $\upsilon_{rd}$ in the form
\begin{equation}
  \upsilon_{rd}=-\frac{\dot M}{2\pi r\Sigma}\,
  \left[1-\left(\frac{r}{R}\right)^{3/2}\right]\,.
\label{eq:11.3.20}
\end{equation}

\noindent Here $\dot M$ is determined by the values at infinity, see Eq.~(\ref{eq:11.3.9}),
$R\approx r_a$ using (\ref{eq:11.3.15}). Ohmic dissipation takes place in the disk,
leading to gas motion through the magnetic field. The surface electrical current density $I_\varphi$ is
determined by an equation
\begin{equation}
  \frac{GM\dot M}{4\pi r^3}\,
  \left[1-\left(\frac{r}{R}\right)^{3/2}\right]=\frac{I_\varphi^2}{4\pi\sigma}\,,
\label{eq:11.3.21}
\end{equation}

\noindent where $\sigma$ is a conductivity. If the disk is opaque in vertical direction to the
radiation, the energy is carried to its surface by a radiative heat conductivity,
so that we have approximately
\begin{equation}
  acT^4=\kappa\Sigma F \,,
\label{eq:11.3.22a}
\end{equation}

\noindent where $T(r)$ is the mean disk temperature. For an optically transparent
disk we have
\begin{equation}
  2F=\Sigma(\varepsilon_{ff}+\varepsilon_B)\,.
\label{eq:11.3.22b}
\end{equation}

\noindent
This relation includes plasma bremsstrahlung $\varepsilon_{ff}$ and
magneto-brems\-strahlung $\varepsilon_B$. Equations (\ref{eq:11.3.16})-(\ref{eq:11.3.22b}) with known
functions $P$, $\kappa$,~$\sigma$,~$\varepsilon_{ff}$ and $\varepsilon_B$ on the average disk temperature $T$ and density $\rho$,  determine
a structure of a non-rotating disk with magnetic field around a BH \cite{bkr76}.

The laminar  disk is always optically thick, electrons are
non-degenerate and non-relativistic, the pressure is determined mainly
by ideal ionized gas $P=\rho {\cal R} T$. The conductivity of non-relativistic electrons, due to Coulomb collisions $\sigma_q$,
neglecting magnetic field dependence, is represented by a simple relation \cite{pik66}

\begin{equation}
\sigma_q= 3\cdot 10^6 \, T^{3/2}\left(\frac{\Lambda}{10}\right)^{-1},
\label{qul}
\end{equation}
where $\Lambda\approx 10$ is the Coulomb logarithm.
Two regions may be separated in such a disk. In an outer region,
bremsstrahlung and photo-ionization processes with \cite{bkr76}
\begin{equation}
\kappa_{ff}+\kappa_{bf}
\approx 2 \times 10^{24}\rho T^{-7/2}\quad {\rm at}\,\,\, kT\ll m_e\, c^2;\qquad
\kappa_{ff}=5\times 10^{18}\, \frac{\rho}{T^3}\quad {\rm at}\,\,\, kT\gg m_e\, c^2,
\label{eq:11.3.22c}
\end{equation}
\noindent
determine the opacity. In the inner one there is a dominance of
the opacity due to magneto-bremsstrahlung absorption

\begin{equation}
\kappa_B \approx 40\, B^2/T^3, \quad {\rm at}\,\,\, kT\ll m_e\, c^2;
\qquad \kappa_B=4.1\cdot 10^{-8} \frac{B^2}{T^2}\quad {\rm at}\,\,\, kT\gg m_e\, c^2.
\label{eq:11.3.22d}
\end{equation}
\noindent
 Since the Coulomb conductivity is large,
the material penetrates slowly through magnetic force lines, and in
the stationary case the disk mass turns out to be large: for a BH of
$M=100\Mo$ the mass of the stationary disk is $M_d\approx 0.2\,\Mo$.
Most of the disk mass is accumulated in its outer parts. The inner disk
is the main source of radiation, the temperature there reaches
$10^8-10^9~{\rm K}$, the magnetic field $10^{10}-10^{12}~{\rm G}$.
The disk thickness does not exceed $\sim 0.01$ of the radius.

In the turbulent disk the dissipation is going much more rapidly
due to meshing of magnetic field lines, and the coefficient of the turbulent
electrical conductivity is approximated by an expression \cite{bkr76}

\begin{equation}
  \sigma\approx\sigma_{turb}\approx\frac{c^2}{\tilde\alpha 4\pi h\sqrt{P/\rho}},
  \quad \tilde\alpha=0.1-0.01\,.
\label{eq:11.3.23}
\end{equation}

\noindent Outer regions of a turbulent disk are transparent for radiation,
electrons are non-relativistic there, the gas pressure predominates,
and contributions of free-free $\varepsilon_{ff}$ and free-bound $\varepsilon_{fb}$ radiation are comparable with magneto-bremsstrahlung losses $\varepsilon_B$ from (\ref{ebnr}). Here we have for non-relativistic plasma

\begin{equation}
\varepsilon_{ff}+\varepsilon_{fb}\approx 2\times 10^{22} \rho T^{1/2}\,,
\label{eq:11.3.23a}
\end{equation}
\noindent
 At ${(M\, r_g / M_{\odot}\, r)}\,=mx\ge 100\tilde\alpha^2$
electrons are relativistic, and $\varepsilon_B$ from Eq.~(\ref{ebr})
greatly exceeds
\begin{equation}
\varepsilon_{ff}^{\rm rel} \approx
2\times 10^{16}\rho T\ln\frac{kT}{m_{\rm e}c^2}\,,
\label{eq:11.3.23b}
\end{equation}
\noindent
see  \cite{bkr76}. We used a connection
between opacity and emissivity  $\varepsilon\,=\,A\sigma T^4 \kappa$, following
from the Kirchhoff's law, in accordance with
the corresponding constant for non-relativistic Rosseland $\kappa_{ff}$ and
$\varepsilon_{ff}$, calculated in textbooks
(see, for example, \cite{bk01}). The same constant $A\,=\,170$ was taken
in estimations of
$\kappa_B$, and for the relativistic case of $\kappa_{ff}$,

 The zone of relativistic electrons is narrow in radius
because the optical thickness increases rapidly and the disk becomes opaque
with decreasing radius. In the disk interiors the gas pressure and electron
scattering opacity,

\begin{equation}
\label{kthoms}
 \kappa_{es} = 0.19\,(1+X_H), \quad X_H \,\,\, {\rm is\,\, the\,\, hydrogen\,\, mass\,\, fraction\,\, in \,\, accreting\,\,  matter},
\end{equation}
predominate for
$10<mx<1000$, while for $mx>1000$ the  pressure is
 radiation dominated. The mass of a turbulent disk is always small because of
a large dissipation and rapid infiltration of gas through the magnetic field lines.
Note the non-monotonic character of the function $T(r)$ connected with transition
of a transparent disk into an opaque one with decreasing $r$.

 The choice between laminar or turbulent disk models could be done after investigation of the stability of the disk. The calculations carried out in \cite{bkb72} show that the  azimuthal currents in the disk tend to stabilize it against fragmentation. Therefore, we cannot exclude here a possibility of a laminar disk. In this sence we find a significant difference between the magnetically supported disk, and the keplerian disk at accretion in binary systems, where differential rotation in combination with magnetic field always produces a turbulence, needed to explain observations.
 Relations describing a distribution of disk parameters
are given in \cite{bkr76} for various disk regions in the turbulent, and laminar disk with a Coulomb conductivity.

For a laminar disk assumptions about the thinness $(h/r \ll 1)$, and quasi-stationarity of the disk with $\upsilon_r \ll \sqrt{2GM/r}$ is fulfilled only if its mass $M_D$ is not too large. The restriction on the disk mass $M_D$ appears because the stationary disk mass is rapidly increasing with a BH mass, M, and for a reasonable $\dot M$, it takes a longer time to establish the quasi-stationary density distribution. Estimations have shown, see \cite{bkr76}, that for $M \, > 100\, M_\odot$ the stationarity is reached after a time, exceeding the age of the universe $\tau_{u}\approx 5\cdot 10^{17}$ c. For a large $M$ it is mainly collecting in the external regions of the accretion disk.

The turbulent disk structure is very different. In the laminar Coulomb disk the temperature continuously increases to the centre, while  in the turbulent disk its distribution is more complicated. In the external parts  the disk is transparent, the energy losses are determined mainly by free-free transitions and the temperature rises towards to the center. Then the non-relativistic magneto-bremsstrahlung processes becomes important and the temperature increases until the electrons become relativistic. Then in the ultra-relativistic condition the temperature in not growing, and if the masses are sufficiently large $(M >10\, M_\odot)$ an internal nontransparent region exists where the electrons are non-relativistic again and the temperature increases to the centre \cite{bkr76}.
For given values of $\rho_\infty$, $T_\infty$, $M$, the mass flux is
determined by Eq.~(\ref{eq:11.3.9}).

\subsection{High energy radiation from the shock on the disk surface, and radiation spectrum}

In the quasi-stationary picture of MAD a matter feeds the accretion disk from two sources. The main gas is coming from extended space in the form of accretion disk, diffusing through the magnetic "wall" due to finite conductivity. In the presence of rotation the gas comes in the form of a Keplerian accretion disk. In a stationary state the poloidal magnetic field almost reaches an equipartition  with the kinetic energy of the falling matter, and gas moves along magnetic field lines in both hemispheres, until it collides with the existed accretion disk, see Fig.~\ref{fig3}. The standing shock front is formed in this collision, and  falling gas joins the accretion disk, increasing the mass flux with approach to a BH. The total mass flux $\dot M$, arrived from infinity, is divided between these two sources. In absence of rotation the disk structure starts to form \cite{bkr76} near the accretion radius $r_a\equiv R$, giving the distributions presented in Eqs.~(\ref{eq:11.3.15}),(\ref{eq:11.3.19})- (\ref{eq:11.3.21}).

The strong deviations from the spherical-symmetric flow, due to large scale magnetic field structure, are essential for formation of a hard component of the radiation in this disk. Indeed, the collision of matter falling  along magnetic field lines on the disk surface leads to formation of a shock wave, in which the kinetic energy of infalling matter is transformed into heat. The shock wave is apparently collision-less, with a thin front \cite{bkf69}; therefore, the temperature of the matter, in presence of ultra-relativistic electrons and non-relativistic protons is determined by the relation

\begin{equation}
 \frac{9}{2}kT=\frac{m_p\, \upsilon^2}{2}.
\label{shd}
\end{equation}
The energy released in the shock is determined by the gravitational potential at the place of a shock formation, and is equal to (see \cite{bkr76})

\begin{equation}
L_{sh}\simeq \int_{1.5r_g}^\infty \frac{d\dot M}{dr}\frac{GM}{r}dr
\label{shd1}
\end{equation}
The dependence $\dot M(r)$ was obtained in \cite{bkr76}, from what it follows, that the hard radiation from the shock $L_{sh}$ contains only a small part of the luminosity, so that for $M_{BH}=10M_\odot$ there is $L_{sh}/L\simeq 10^{-4}$. The hard radiation spectrum from the shock is estimated in \cite{bkr76} for $M_{BH}=10 M_\odot$. It has an almost power law, with an exponential cutoff at frequency $\omega\approx 3\cdot 10^{23}\,$s$^{-1}$, corresponding to the energy about 500 MeV.

As the gravitational energy always has time to convert into heat during
stationary disk accretion, the black-hole luminosity for a minimum radius
of $\sim 1.5\,r_g$ is
\begin{equation}
  L\approx \frac{GM{\dot M}}{r_g}=\frac{1}{2}\,\dot Mc^2=5\cdot 10^{31}\MM2\rhoinf
 \times\TK{-3/2}~{\rm erg~s^{-1}}
\label{eq:11.3.24}
\end{equation}

\noindent with $\dot M$ from Eq.~(\ref{eq:11.3.9}), determining very high efficiency of this accretion regime, which  was recently  named as magnetically arrested disk (MAD) \cite{nia03}.
The emission spectrum of an opaque disk is related to its
effective temperature determined locally by
\begin{equation}
  \frac{ac}{4}\,T\r{ef}^4=F
\label{eq:11.3.25}
\end{equation}

\noindent when
the shock energy is supposed to be completely thermalized. Hence, using
Eq.~(\ref{eq:11.3.19}), we
have for the inner parts of the disk around a BH with $r\ll R$, where the most part of the energy is
released,
\begin{equation}
  T\r{ef}\sim r^{-3/4}\,.
\label{eq:11.3.26}
\end{equation}

\noindent After integration over the disk surface with the temperature distribution (\ref{eq:11.3.26}), we obtain the spectral distribution

\begin{equation}
  L(\omega)\sim \omega^{1/3},
\label{eq:11.3.27}
\end{equation}

\noindent with an exponential cut-off in the soft $X$-ray region, at $\hbar\omega\sim kT\r{max}$,
$T\r{max}\approx 7\times 10^5~{\rm K}$ for $M=10\Mo$. So, despite a
substantial differences in flow patterns in cases of a random and ordered magnetic
fields, the spectrum of the optically thick disk is similar to the random field case,
and the relation (\ref{eq:11.3.14}) for $m_\upsilon$ is valid in such a case as well.
A turbulent disk has a large transparent region with magneto-bremsstrahlung
in the infrared range that may be comparable in power with the ultraviolet
and soft X-ray emission of opaque disk interiors. In the case of accretion with a large scale magnetic
field  a small, almost power-law  hard tail is present, continuing until $\sim 500$ MeV. Such a tail is absent in the case of a spherical accretion with a fully chaotic magnetic field.

If the accreting gas has an intrinsic angular momentum, the magnetized
disk generates electric fields \cite{bkb72,bkb76} of the strength
$E \approx -(\upsilon/c)B$ . In this field the electrons are accelerated up to energies
$\varepsilon \approx R(\upsilon/c)Be \approx 3\cdot 10^4 [B/(10^7\,\mbox{Gauss})]$ Mev
where $\upsilon/c \approx 0.1$ and $R \approx 10^7$ cm
 is the characteristic scale. In a field $B \approx 10^7$ Gauss,
such electrons generate synchrotron radiation with
energies up to $\approx 10^5$ keV, which flux may be much larger than the
hard radiation flux from the shock.
 Like in pulsars, it would be
possible here for $e^{+}e^{-}$  pairs formation, which
participate in the synchrotron radiation.
Such a mechanism is likely to act in Cyg X-1, galactic nuclei
\cite{lov76,bz77}, and is analogous to the unipolar mechanism proposed for explaining
the pulsar radiation \cite{gj69}.

Using the MAD solution the gas rotation is small in the inner part of the accretion disk. In this case a
formation  of the electrical field may be connected with  BH rotation, and processes inside the ergosphere,
where matter rotation becomes inevitable. This condition was considered in \cite{bz77}, for the extraction
of the rotational energy from the BH.

\section{Numerical simulations of magnetically arrested disk}

Two- and three-dimensional MHD simulations had been performed in \cite{ig08}
for investigation of the dynamics
and structure of magnetically arrested disk (MAD), in presence of a rotation. It was obtained.
that MAD was formed in accretion
flows, carrying inward large-scale poloidal magnetic
fields. Because of
rotation, the streams take spiral shapes.
 The jets formation takes place in MAD,
because of the interaction of the spiraling accretion flow with the
vertical magnetic bundles, twisted around the
axis of rotation.
 Ideal non-stationary MHD 3-D equations have been used with an ideal gas equation of state

 \begin{equation}
 P= (\gamma-1)\rho\varepsilon,
\label{mad1}
\end{equation}
 at adiabatic power $\gamma=5/3$. Energy equation was included in the system because a growth of the entropy due to numerical resistivity and
viscosity, and in  possible MHD shocks. Self-gravity of
gas was neglected, and a pseudo-Newtonian approximation \cite{pw80}  was employed for the BH potential.

 \begin{equation}
 \Phi=-\frac{GM}{R-R_g},\qquad R_g=\frac{2GM}{c^2}\,\,\,{\rm is\,\,\, the\,\,\, BH\,\,\, gravitational\,\,\, radius}.
\label{mad2}
\end{equation}
Radiatively inefficient accretion disk
with poloidal magnetic fields was considered, without a cooling term in the energy equation, similar to  \cite{ina03}, see however  \cite{bkl97}.
 The MHD equations have been solved by using the time explicit
Eulerian finite-difference scheme, which is an extension
to MHD of the hydrodynamic piecewise-parabolic method by \cite{cw84}. Spherical coordinates $(R,\,\theta,\, \phi)$ were used. The 3D numerical grid
had $182 \times 84 \times 240$ zones in the radial, polar, and azimuthal directions,
respectively. The radial zones were spaced logarithmically
from $R_{in} = 2\, R_g$ to $R_{out} = 220\,R_g$. Two polar cones with an opening
angle of $\pi/8$ were excluded. So, the polar domain extends
from $\theta = \pi/16$ to $15\pi/16$. The grid resolution in the polar
direction was gradually changed from a fine resolution around the
equatorial plane to a coarse resolution near the poles, with a
maximum-to-minimum grid size ratio $\approx 3$. The azimuthal zones
 cover uniformly  the full $2\pi$ range in $\phi$,
  see more detail of the numerical method in \cite{ig08}.
Initially, the computational region
was filled with a very low-density, non-magnetized gas. The simulations
were started in 2D, assuming axial symmetry, with an injection
of mass with a finite entropy, and Keplerian angular momentum in a slender torus, located in the equatorial
plane at $R_{inj} = 210\,R_g$. Simulations without the magnetic field were finished
after formation of a steady thick torus, with
the inner edge at $R_{in}\approx 150\,R_g$, and the outer part is truncated
at $R_{out}$. The torus contained a constant amount of mass and is in dynamic
equilibrium: all injected mass flows outward through
$R_{out}$ after circulation inside the torus. No accretion flow was formed
at this point. This hydrodynamic, steady, thick torus was used as an
initial configuration for the MHD simulations.

\begin{figure}[t]
\centerline{\includegraphics[width=7.0cm]{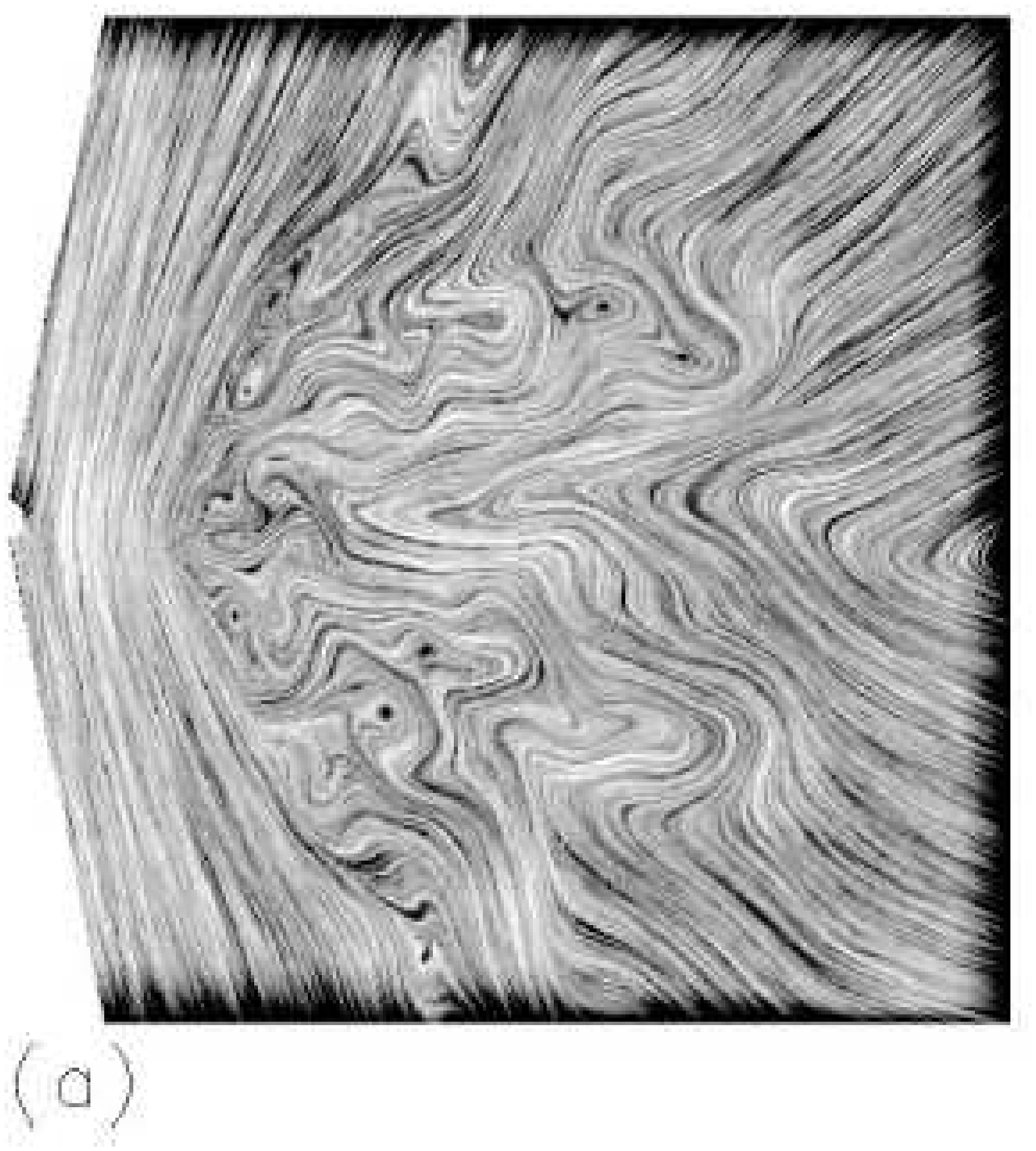}
           \includegraphics[width=7.5cm]{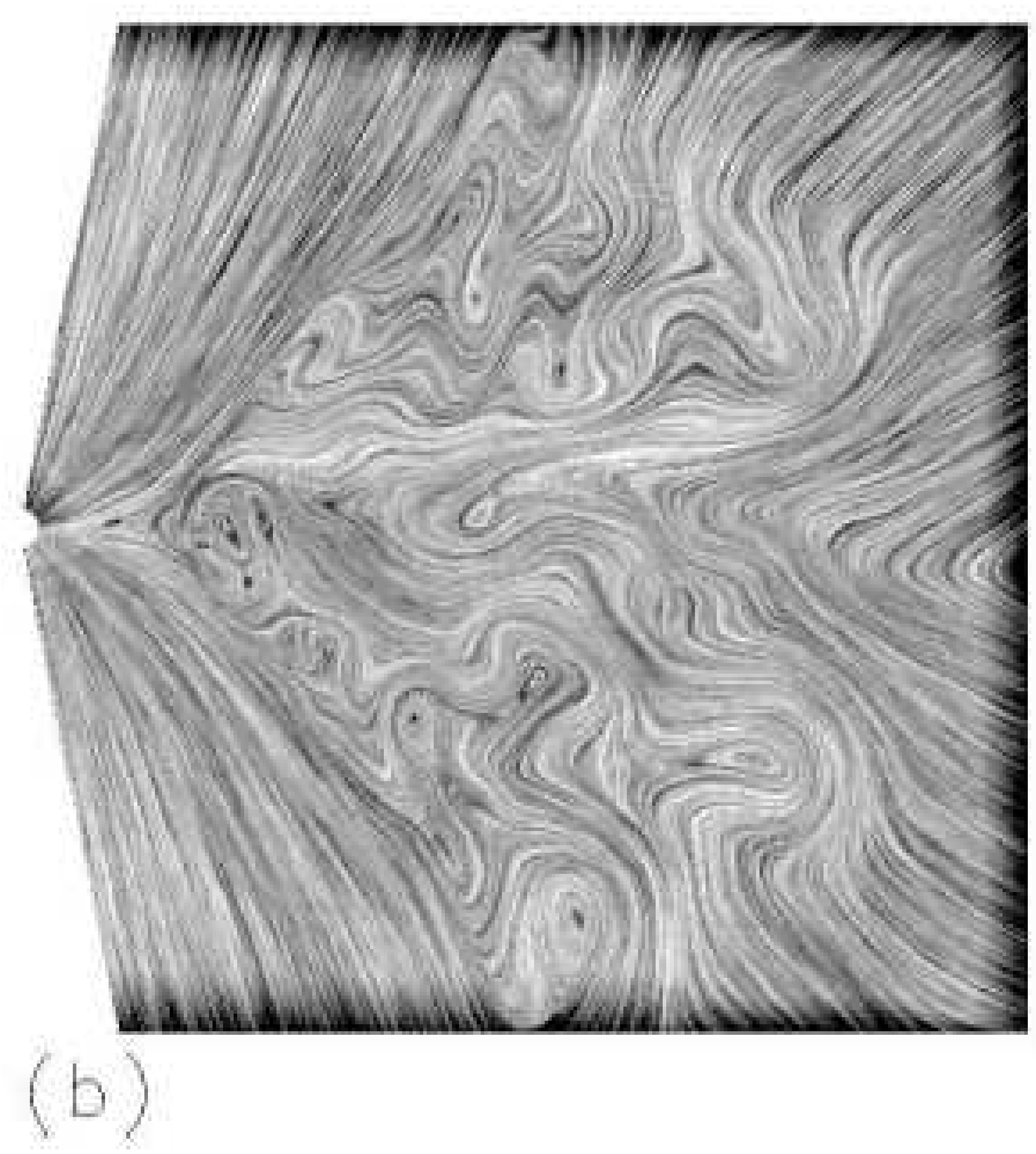}}
\caption{Snapshot of magnetic lines in model with $\beta_{ing}=100$,  at two subsequent moments.
The BH is located on the left, and small open circle corresponds to the inner boundary around the black hole at $R_{in} = 2\,R_g$. The axis of rotation is in the vertical
direction. The domain in the figure has a radial size $100\, R_g$ along the equatorial plane and represents a fraction of the full computational domain with $R_{out} = 220\,R_g$.
 The poloidal field lines lying in the
meridional plane are shown. The accretion disk transports the vertical magnetic flux inward, which is accumulated in the vicinity of a BH. Small-scale magnetic loops
are the result of turbulent motions in the disk and disk corona. (a) Period of accretion, in which most of the accumulated magnetic flux is outside the
black hole horizon. (b) Accretion period, in which all the accumulated flux goes through the horizon,
\cite{ig08}.}
\label{fig4}
\end{figure}

The MHD simulations were started at $t = 0$ from the steady,
thick torus by starting injection of a poloidal magnetic field
into the slender torus at $R_{inj}$. The numerical procedure for the
field injection is described in \cite{ina03}.
The entire volume of the thick torus is filled by the field during
about one orbital period, $t_{orb}$, estimated at $R_{inj}$. Starting from the moment,
$t \simeq t_{orb}$, formation of the accretion flow begins as a result of
redistribution of the angular momentum in the torus due to
magnetic forces. Three models were calculated with $\beta_{ing}=10,\,100,\,1000$, which is
the ratio of the gas pressure to the magnetic pressure at $R_{inj}$. After beginning of the accretion
an evolution of disks
is governed mainly by magnetic forces produced by the
poloidal field component. This component is advected inward
with the accretion flow and
generates strong toroidal magnetic fields, because of the Keplerian rotation, localized
above and below the midplane.
These toroidal fields form a highly
magnetized disk corona with a typical $\beta\sim 0.01$.
Due to the permanent mass injection
 and magnetic field
accumulation of a poloidal field happens in the central parts of the accretion disk where
MAD is formed. The computational results for $\beta_{ing}=100$ from \cite{ig08} are given in Fig.\ref{fig4}. MAD formation in the central parts is clearly visible in this Figure.
Note, that in the quasi-Newtonian approximation magnetic flux after finite time goes through the horizon. In the GR consideration the distant observer will not see the penetration of the magnetic flux through the horizon, because it take for him an infinite time to reach the horizon for the magnetic flux, as well as for the matter \cite{zn71}. It does not change the conclusion about a very high efficiency of MAD for producing  radiation flux, which is coming from the accretion disk itself. at the finite distance from the horizon.

3-D finite conductivity MHD simulations of the cooling instability
in optically thin hot BH accretion flows had been studied in \cite{mnm06} assuming the pseudo-Newtonian gravitational potential \cite{pw80}, and bremsstrahlung cooling.  It was found that the cooling instability  changed the accretion flow from an optically
thin, hot, gas pressure-supported state to a cooler, quasi-steady state MAD structure.
 The stability of the MAD disk was examined analytically, and it was found to be thermally and secularly stable.
3D simulations of accreting BHs had been performed in \cite{tnm11}, using time-dependent, non-radiative, general-relativistic, magnetohydrodynamic equations.
The main difference from \cite{ig08} was using full GR equations instead of the pseudo-Newtonian potential \cite{pw80}.
In these simulations
a large amount of magnetic flux was transported to the centre, remains outside a BH, impedes accretion, and leads to a magnetically arrested disk.  Powerful outflows have been found, similar to \cite{ig08}.
An efficiency $\eta$ of transformation of the gravitation energy of accreted mater into the
out-flowing energy in jets and winds,
depended on a BH spin parameter in the Kerr metrics $a$. It changed from  $\eta \approx 30\%$  for  $a = 0.5$, up to $\eta \approx 140\%$ for $a = 0.99$, what is an indication to additional energy supply from the rotational energy of BH.
 The energy extraction from a spinning BH could happen only via the Penrose mechanism \cite{pen69}, generalized for the presence of a magnetic field in \cite{rw75,bz77}.

 Equilibrium of  two-temperature optically thin MAD disk was studied in \cite{omnm10} using equations averaged over the thickness, with pseudo-Newtonian BH gravitational potential \cite{pw80}. Outflows had not been included into consideration.
  The first Magnetically Arrested Disc simulations  with two-temperature,
radiative, general relativistic magnetohydrodynamics were performed in \cite{cnj18} in 3-D modeling. Mass accretion rates of the simulations were scaled to match the luminosity of the
accretion flow around the super-massive black hole in M87. Several mechanisms of electron heating were considered. The main features of the numerical solution resemble that produce in \cite{ig08,tnm11}.

The stationary picture of the accretion disk around a BH is formed due to violation of the ideal MHD condition, and finite conductivity of a disk matter. The magnetic flux growth stops, when rate of matter penetration trough the magnetic field lines balance the accretion gas flux from large distances. The particular structure of the MAD depends strongly on electrical conductivity of the disk matter. Laminar Coulomb and turbulent electrical conductivities were investigated in \cite{bkr76}, see Section 4. It was shown, that only phenomenological turbulent conductivity coefficient, derived in \cite{bkr76}, gives a physically relevant stationary solution of MAD. Numerical simulations using ideal MHD \cite{ig08,tnm11,cnj18}, may be used only for limited period of time, and simulations with a finite electrical conductivity \cite{mnm06} permitted to obtain a physically realistic stationary solution.

\section{Magnetized disk levitation and MAD}

Many interesting effects appear in consideration of the MAD around neutron stars. In this case the MAD is restricted by the star surface, or by the Alfven radius in magnetized stars, where the pressure of its magnetic field is equal to the large scale magnetic field pressure in the disk. This model was considered in several papers with an application to explain observational properties of different objects. It was called as  magnetic-levitation accretion scenario \cite{ikkb15}. Consider several objects where this model was applied for modeling the observational data.

\smallskip

  {\bf 1}. The spin-down mechanism of accreting neutron stars was discussed in \cite{ikb12,ikf12}, with an application to the  X-ray pulsar GX301–2, showing rotation period $P_s=685$ s. It was obtained that the maximum possible spin-down torque applied to a NS from the accretion flow can be evaluated as $K^{(t)}_{sd}=\mu^2/(r_m\, r_{cor})^{3/2}$.  Here $r_{cor} =(GM_{ns}/\omega_s^2)^{1/3}$ is
the co-rotation radius of a NS  spinning
at angular velocity $\omega_s = 2\pi/P_s$, $\mu$ is its dipole magnetic moment, $r_m$ is the
radius of the neutron star magnetosphere (Alfven radius). The spin-down rate
of a NS in GX301–2 can be explained if the magnetosphere radius of the neutron star
is smaller than its canonical value.  The magnetosphere radius was calculated in \cite{ikb12} considering the mass-transfer
in the binary system in the model of the magnetic-levitation accretion scenario. It was shown, that the
spin-down rate of a NS expected within this approach is in a good agreement with that derived from observations of GX301–2. Other pulsars of this type were considered in \cite{iklb14,ikbl14}.

\smallskip

   {\bf 2}. Observations of the X-ray pulsar 4U 2206+54 during more than 15 years have shown that its period,
which is now $5555 \pm 9$ s, is rapidly increasing. Such a behavior is difficult to explain inside traditional scenarios for the spin evolution of compact stars. It was shown \cite{ikb13}, that the observed spin-down rate of a NS in 4U 2206+54 is in a good agreement with the value expected in a magnetic-accretion scenario, if the magnetic field of the accretion stream  affects the geometry and type of flow. A NS in this case accretes material from a dense gaseous slab with a small angular momentum, which is kept in equilibrium by the magnetic field of the flow. A magnetic accretion scenario can be realized in 4U 2206+54 if the magnetic field strength at the surface of the optical counterpart to a NS is higher than 70 G. The magnetic field at the surface of the neutron star  in this scenario is  $\sim 4 \cdot 10^{12}$ G, in agreement with estimates based on an analysis of X-ray spectra of the pulsar.

\smallskip

  {\bf 3}. Growth of the NS period of in the young Be/X-ray long-period
pulsar SXP 1062 was considered in \cite{ik12}. The observed period is about 1062 s, and the star is spinning down at the rate $\sim - 2.6 \cdot 10^{-12}$ Hz/s. It was shown that all of the conventional
accretion scenarios meet big difficulties in explaining the rapid spin-down of this pulsar. It was shown in \cite{bkr74,bkr76,ina03}, that these difficulties can be avoided within the magnetic accretion scenario, where a NS is assumed to accrete from a magnetized wind. The spin-down rate of the pulsar can be explained
within this scenario provided the NS surface magnetic field is $B_* \sim 4 \cdot 10^{13}$ G. It happens, that the age of the pulsar in this case lies in the range $(2 - 4) \cdot 10^4$ yr, which is consistent with observations.

\smallskip

  {\bf 4.} A scenario of the formation of isolated X-ray pulsars was discussed in \cite{ikkb15,ikkbp13}  with an application to the object 1E 161348-5055. This moderately luminous, $10^{33} - 10^{35}$ erg/s
pulsar with a relatively soft spectrum, $kT \sim 0.6 - 0.8$ keV, is associated with an isolated neutron star, which is located near the center of the young ($\sim 2000$ yr) compact supernova remnant RCW 103, and rotates with a period of 6.7 h, and slowly decreasing frequency ($|\dot \nu | \le 2.6 \cdot 10^{-18}$ Hz/s). It was shown that at the present epoch the NS is in the accretor state. The parameters of the source emission can be explained in terms of the magnetic-levitation accretion scenario in which a star with the surface magnetic field of $10^{12}$ G accretes matter onto its surface from a non-Keplerian magnetic fossil disk, at the rate $10^{14}$ g/s. A neutron star could evolve to this state in a High-Mass X-ray Binary (HMXB), which had disintegrated during the supernova explosion powered by the core-collapse of its massive component. A life-time of an isolated X-ray pulsar, formed by this way, can be of about few thousand years.

\smallskip

  {\bf 5.} A behavior of Anomalous X-ray Pulsars (AXPs) and Soft Gamma ray
Repeaters (SGRs) was considered in \cite{bki14}, in a scenario with fall-back magnetic accretion onto a young isolated neutron star.
The X-ray emission of the pulsar in this case originates due to the accretion of matter onto the surface of a NS from a magnetic slab surrounding its magnetosphere. The spin-down rate of a NS is expected in this picture to be close to the observed value. Such NSs should be relatively young. The pulsar’s activity in gamma-rays is related to its relative youth, and is enabled by energy stored in a non-equilibrium layer \cite{bkch74,bkch79},
located in the crust of a low-mass NS. Outbursts are probably triggered by instability developing in the region where the accreted matter is accumulated.

\smallskip

There are also applications of the MAD model to different astronomical objects, containing a BH.

\smallskip

{\bf 1}.
Simulations of the axisymmetric MAD disk - jet system in the pseudo-Newtonian gravitational potential \cite{pw80}, and an approximate treatment of the derivative over $z$ - coordinate, with account of different heating and cooling mechanisms in the disk and jet, had been done in \cite{mm19a}. The magneto-centrifugally driven outflows from the disk threaded by the large scale open magnetic field was obtained.
The results have been used for a unified classification of blazars.

\smallskip

{\bf 2}.
2.5-D simulations of the geometrically thick, sub-Keplerian, magnetized, viscous, advective, coupled disc-jet model have been done in \cite{mm19}. A gravitational force was obtained using the pseudo-Newtonian potential. It is claimed that the maximum possible outflow power in this model is equal to  $7.5 \cdot 10^{39}$ erg/s for a non-rotating stellar mass ($20\, M_\odot)$)  black hole accreting at sub-Eddington accretion rate. The outflow power extracted from the disc is defined by the combination of mechanical, enthalpy, viscous and the Poynting parts. Hence, this scenario could provides an explanation of the  nature of hard-state ULXs without introducing  the existence of the missing class of intermediate mass BHs, nor with the super-Eddington accretion \cite{mm19}.

\smallskip

{\bf 3}.
 The main purpose of the paper \cite{omnm10} was to explain the bright/hard state observed during the bright/slow transition in the rising
phases of transient outbursts of galactic BH candidates, like Cyg X-1.
 In the low/hard state,
the X-ray spectrum is described by a hard power law with a
high energy cutoff at $\sim 200$ keV. With increasing luminosity, the source enters the bright/hard state, when the cutoff energy decreases down to $\sim 50$ keV. This transition was explained in \cite{omnm10} by formation of MAD, as a result of increasing of the accretion rate.

\smallskip

{\bf 4}.
It is shown in \cite{llx18} that a variability timescale in the MAD
model as defined in \cite{ldf16} can reproduce
the observed minimum variability timescale (MTS) -
bulk Lorentz factor correlation as well as the MTS - luminosity
(L) relation observed in the long GRB data, and can
also be applied/extended to AGN data.

\section{Conclusions}

During accretion of the matter with a large scale magnetic field into a BH the disk is formed around a black hole, which equilibrium is supported by the balance between a BH gravity and a magnetic pressure. The semi-analytical models of such disk a have been constructed in 70s \cite{bkr74,bkr76}. An efficiency of transformation of the gravitational energy into the radiation in this model is very high, with $L/\dot M\, c^2\, \sim 0.5$. Such a high efficiency is connected with an action of the external field, which holds the disk from falling into BH up to smaller radius $\sim 1.5 r_g$. This model of accretion attracted attention in 2003 \cite{nia03}, where it was called as a magnetically arrested disk (MAD), reflecting a role of the magnetic field, holding the matter from falling into a BH. Numerical simulations of the MAD model confirmed the semi-analytical results \cite{bkr74,bkr76,nia03}, being done in the framework of a pseudo-newtonian gravitational potential \cite{pw80,ig08}, as well as in the full GR consideration \cite{tnm11,cnj18}. In an application of this model to accretion of the magnetized matter onto a NS it was called as magnetic-levitation accretion scenario \cite{ikkb15}. Application of this model for an explanation of observational properties of X-pulsars of different types, and behavior of objects, containing a BH, occurs to be rather promising.

\section*{Acknowledgments}
This work was supported by Russian Science Foundation, Grant No. 18-12-00378.





\end{document}